\documentclass[proof]{WileyASNA-v2}
\usepackage{natbib}
\articletype{Article Type}%

\received{26 April 2019}
\revised{10 July 2019}
\accepted{15 July 2019}

\begin{document}

\title{Tidal disruption events as a site of an evolving relativistic spectral~line}

\author[1]{Marcel \v{S}tolc*}
\author[2]{Vladim\'{\i}r Karas}
\address[1]{\orgdiv{Astronomical Institute, Faculty of Mathematics and Physics}, \orgname{Charles University}, \orgaddress{\state{V Hole\v{s}ovi\v{c}k\'ach 2, CZ-18000 Prague}, \country{Czech Republic}; \email{stolcml@gmail.com}}}
\address[2]{\orgdiv{Astronomical Institute}, \orgname{Czech Academy of Sciences}, \orgaddress{\state{Bo\v{c}n\'{\i} II 1401, CZ-14100 Prague}, \country{Czech Republic}; \email{vladimir.karas@cuni.cz}}}
\corres{* Bo\v{c}n\'{\i} II 1401, CZ-14100 Prague, Czech Republic. This contribution has been based on a student project by M.\,\v{S}. at the Astronomical Institute of the Charles University in Prague.}

\abstract{In nuclei of galaxies strong tidal forces can destroy stars passing within a critical distance from the central super-massive black hole (SMBH). Observational signatures of tidal disruption events (TDEs) depend on the environment around the SBMH horizon and the level of its accretion activity. Evidence for optical and UV spectral features has been reported in TDE flares; nevertheless, to test the effects of General Relativity in the immediate vicinity of SMBH, the relativistically broadened and skewed X-ray line would tell us significantly more useful information. This will require to proceed beyond inactive nuclei. To this end we consider a system where the material from a disrupted star forms a gaseous ring that circularises near the tidal radius around SMBH, it gradually spreads in radius by viscous processes, and resides embedded within a hot corona. 

In our test calculation the remnant trail is assumed to be fully circularized and embedded in a hot environment and illuminated by X-rays from a surrounding corona or a jet base of (mild) AGN activity. We show the expected effects on the observed profile and the centroid energy. In future the evolving spectral features can enhance the diagnostic capability and provide a novel way to reveal the parameters of TDEs in such sources; namely, the distance of the remnant gas from the SMBH, the radial extent of the gaseous trail, and the spin of the SMBH could be measured.}

\authormark{M. \v{S}tolc and V. Karas}
\jnlcitation{\cname{%
\author{\v{S}tolc, M., and Karas, V.}} (\cyear{2019}), 
\ctitle{Tidal disruption events as a site of an evolving relativistic spectral~line}, \cjournal{Astronomische Nachrichten}, \cvol{2019;}.}
\maketitle
\let\thefootnote\relax\footnotetext{Abbreviations: AGN, active galactic nucleus; ISCO, innermost stable circular orbit; SMBH, super-massive black hole; TDE, tidal disruption event.}

\section{Introduction}\label{section1}
The potential importance of Tidal Disruption Events (TDEs) as a way to probe the environment of supermassive black holes was recognized in the early 1980s \citep{hills,1985MNRAS.212...57L,rees}. It has been soon recognized that the mechanism of circularization and the onset of accretion play an important role in the way how TDEs can be revealed observationally \citep{1989ApJ...346L..13E}. Since then, over almost three decades, the understanding of the mechanism of TDEs and different flavours of the process affecting the plunging stars was examined, first mainly theoretically and later on also observationally \citep{Komossa,Stone,2019GReGr..51...30S}. Furthermore, it has been argued and the evidence confirms that TDEs can provide new constraints of dormant supermassive black holes in normal galaxies \citep{2004ApJ...603L..17K,2012A&A...541A.106S}. These are governed by different time-scales. During the rising phase, the accretion luminosity increases by orders of magnitude in short time (order of weeks), and the emergent ionizing radiation illuminates the fresh accretion flow. However, in active galactic nuclei (AGN) the detection of TDEs has been hampered by the overall accretion variability in the gaseous flow, especially in the X-ray band; nonetheless, ideas were proposed for the prospects of TDE detection in mild AGNs. The evidence of spectral signatures from such events could test both the TDE mechanism as well as the underlying AGN activity \citep{1996rftu.proc..471M,2002JKAS...35..123C,2014bhns.work..129K}.

Once the star gets too close to the SMBH the tidal forces acting upon it overcome its self-gravity and create a gaseous trail of the remnant material residing near the tidal radius,
\begin{equation}\label{tidal_radius}
R_{\rm{tidal}}=\bigg( \frac{2M_{\rm{SMBH}}}{M_{\ast}}\bigg)^{\frac{1}{3}}R_{\ast},
\end{equation}
where $M_{\rm{BH}}$ is the mass of SMBH, $M_{\ast}$ and $R_{\ast}$ are the mass and the radius of passing star respectively.
The newly emerging gaseous trail initially has low angular momentum and only later it circularizes into a fluid ring that spreads by the viscous forces (see Sect.\ \ref{section2}). Given the scaling relations for the characteristic masses, $R_{\rm{tidal}}\propto M_{\rm{SMBH}}^{{1}/{3}}$ for tidal radius and $R_{\rm{g}}\propto M_{\rm{SMBH}}$ for the gravitational radius, respectively, there is an upper limit constraining the SMBH capable of tidally damaging the star: Hills' critical mass is determined by the relation \citep{Stone}
\begin{equation}\label{Hills_mass}
M_{\rm{Hills}}\simeq1.1 \times 10^{8} M_{\odot}\bigg( \frac{R_{\ast}}{R_{\odot}}\bigg)^{\frac{3}{2}} \bigg( \frac{M_{\ast}}{M_{\odot}}\bigg)^{-\frac{1}{2}},
\end{equation}
where $M_{\odot}$ and $R_{\odot}$ are the mass and the radius of sun, respectively.

TDEs vary in different ways over the timescale of the process, but their typical signatures are known well \citep[see][and references therein]{Komossa}:
(i)~the bolometric luminosity decreases $\propto t^{-5/3}$,
(ii)~the soft X-ray spectra hardens as a function of time,
(iii)~no activity of the galactic nucleus observed prior to the TDE,
(iv)~the bolometric luminosities ranges from $\sim10^{45}$--$10^{46}\;$erg/s, peaking in the interval of 0.2--2 keV \citep{monte}. 

The resulting radiation intensity fluctuations typically exceed those expected in classical AGN accretion spectra. Our present work describes the current results from an ongoing student project, where we anticipate the detections of the relativistic spectral features related to TDEs. \citet{magorian} suggest and further evidence confirm that there is a SMBH in the centre of (almost) every galaxy. Also the Nuclear Star Clusters are present in many galaxies, and so TDEs are expected to inevitably occur. Various pieces of observational evidence indicate an emerging jet \citep{cummings} or the presence of spectral features in UV and X-rays (including Fe K$\alpha$ emission) in the highly ionized state \citep{2015ApJ...807...89Z,kara,2018ApJ...859L..20D} during the early stages of TDE.\footnote{Specifically, \citet{2015ApJ...807...89Z} discuss the spectral lines that may appear in the early stages of TDEs.} Furthermore, accretion-driven outflows and winds can be powered by TDEs and lead to absorption features in X-rays that should constrain abundances and velocity fields of the outflowing material near SMBH \citep{2017AN....338..256K}, and TDEs might prove to be a useful tool to explore also the demographics of SMBH. 

Let us note that the above-mentioned characteristics may not be entirely universal in all circumstances; in particular they depend on the the disrupted star type \citep{2018ApJ...857..109G}. It has been argued \citep{2019arXiv190703034C} that the fall-back rate and the implied luminosity decay become slower ($\propto t^{-9/4}$) in the case of only partial disruption when the stellar core survives the pericenter passage and continues to exert the gravitational influence on the stream of tidally-stripped debris. Ignoring these complexities, however, here we adopt the classical (analytical) value of the index $-5/3$ and we focus ourselves on modeling the spectral line profile time evolution of the radiation reflected by the evolving accretion disc, originating as a result of an evolving X-ray illuminated remnant fluid ring that presumably arises from a complete TDE.

\section{Viscous spreading of an accretion ring}\label{section2} 
Accretion is a process mostly studied in the context of binary stellar systems consisting of a donor and an accretor component. For our purposes we will study the behaviour of systems with a SMBH as the accretor and substitute the donor component for a material that has been tidally stripped off of a passing star. We presume the axial (cylindrical) symmetry of the problem which is a simplification allowing us to proceed with the analytical approaches further albeit the realistic astrophysical regimes are fully described by three-dimensional spatial and temporal dependency. The key equations necessary to describe the accretion disc are the continuity equation
\begin{equation}\label{continuity_equation}
R\frac{\partial \Sigma}{\partial t}+\frac{\partial}{\partial R}(R \Sigma v_{R})=0,
\end{equation}
with $\Sigma$ as surface density profile of the accretion disc (here within the Newtonian framework), $v_{R}$ the radial component of a circulating material and angular momentum and energy transfer equation \citep{frank,2008bhad.book.....K}
\begin{equation}\label{momentum_conservation_equation}
R \frac{\partial}{\partial t}(R\Sigma v_{\varphi})+\frac{\partial}{\partial R}(R^{2} \Sigma v_{R} v_{\varphi})=\frac{1}{2 \pi} \frac{\partial G}{\partial R},
\end{equation} 
where $v_{\varphi}=R\Omega$ as $\Omega$ are angular velocity and $G$ is $R$-dependent gravitational torque. The latter describes the angular momentum transfer between neighbouring annuli of the accretion disc.

\begin{figure}[tbh!]
\centerline{\includegraphics [width=0.49\textwidth]{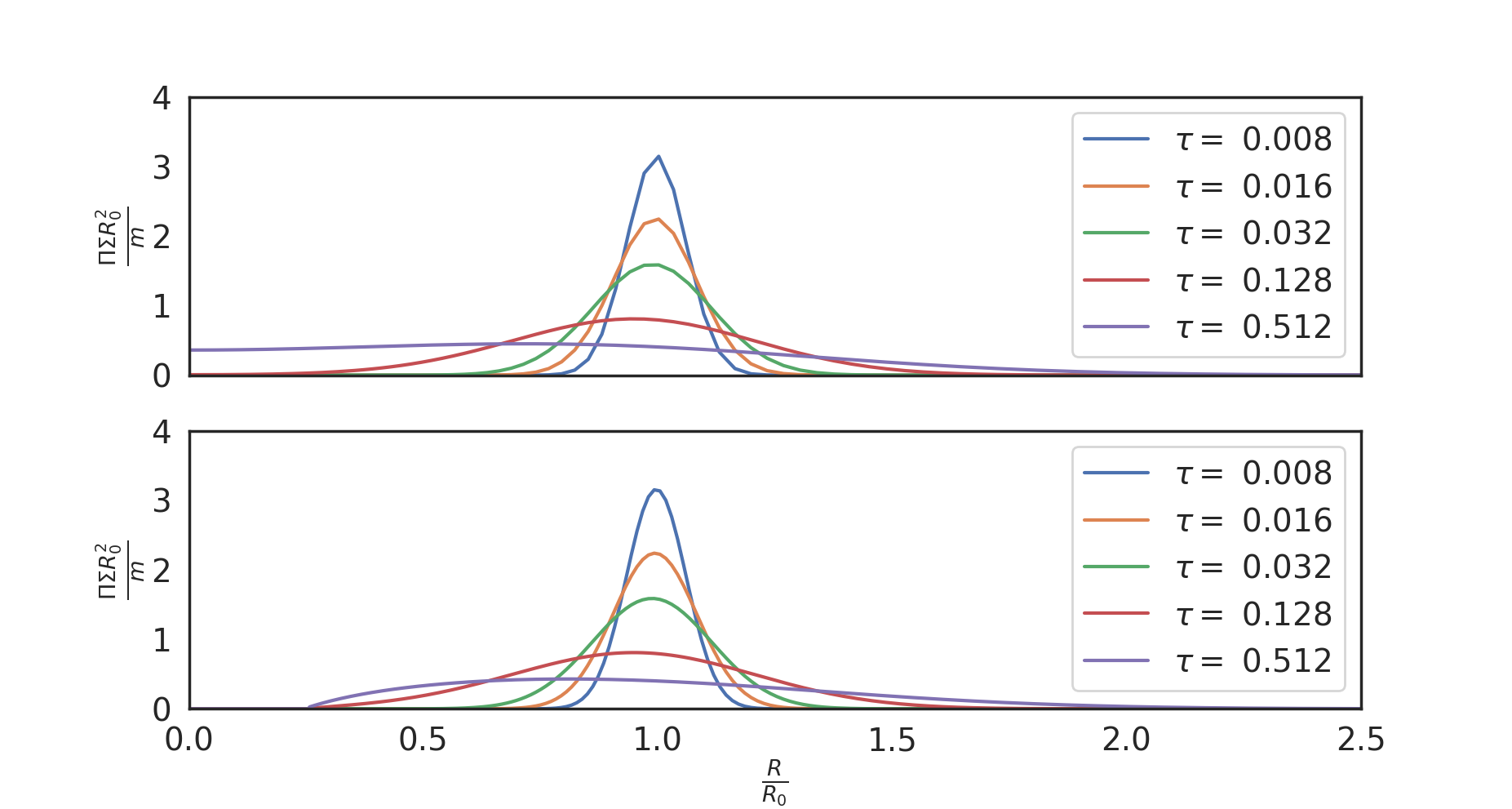}}
\caption{The radial surface density profile for two cases of boundary conditions, as implied by the standard $\alpha$-viscosity prescription for a thin accretion disc: (i) $\Sigma(R_{\rm{inner}}= 0,t) = \Sigma (R_{\rm{outer}},t)= 0$ (top panel), and (ii) $\Sigma(R_{\rm{ISCO}}=6R_{\rm{g}},t) = \Sigma (R_{\rm{outer}},t) = 0$ (bottom panel), with $R_{0}=23.6 R_{\rm{g}}$, where time steps $\tau$ are measured in arbitrary (geometrized) units. The effect of gradual spreading of the matter distribution is seen above the outer and below the inner edges of an initially narrow ring. Difference in the inner boundary condition influences the density profile near the inner rim of the accretion flow, which then affects the expected shape of the spectral features close to SMBH .}\label{boundary_conditions}
\end{figure}

The assumption leading to the eq.\ (\ref{continuity_equation}) and (\ref{momentum_conservation_equation}) are the Newtonian approximation together with the velocity inequalities $v_{\varphi}\gg v_{R} \gg v_{z}$. Combining both of the equations (\ref{continuity_equation}) and (\ref{momentum_conservation_equation}) and taking the adjustments \citep{frank} one arrives at formula
\begin{equation}\label{surfce_density_profile_equation}
\frac{\partial \Sigma}{\partial t}=\frac{3}{R}\frac{\partial}{\partial R}\left[R^{\frac{1}{2}}\frac{\partial}{\partial R}(\nu \Sigma R) \right],
\end{equation} 
which is a partial differential equation describing the spatial and temporal evolution of surface density profile of the accretion disc. A standard form of the solution to the equation (\ref{surfce_density_profile_equation}), taking into account the initial mass ring represented by the Dirac delta distribution $\delta (R-R_{0})$, can be expressed as \citep{frank}
\begin{equation}\label{analytical_solution}
\Sigma(x,\tau)=\frac{m}{\pi R_{0}^{2}}\tau^{-1}x^{-\frac{1}{4}}\exp \bigg\{\! -\frac{(1+x^{2})}{\tau} \bigg \} I_{\frac{1}{4}}\bigg(\frac{2x}{\tau}\bigg),
\end{equation}
with dimensionless quantities $x=\frac{R}{R_{\rm{0}}}, \tau=12 \nu t R_{0}^{-2}$ and modified Bessel functions $I_{\frac{1}{4}}$, whereas $R_{0}$ stands for the initial radial location of the mass ring, $m$ for the mass of the initial mass ring and the kinematic viscosity $\nu$ is assumed to be constant. We can evaluate the evolving surface density, and thus estimate the changing area that forms the reflection line.

The initial step in solving eq.\ (\ref{surfce_density_profile_equation}) numerically is to rewrite its convenient form,
\begin{equation}\label{prepis_surface_density_profile_equation}
\frac{\partial U}{\partial t}=\frac{12 \nu}{x^{2}}\frac{\partial^{2} U}{\partial x^{2}},
\end{equation}
and substitute $U=\Sigma R^{\frac{1}{2}}$ and $x=2R^{\frac{1}{2}}$ \citep{frank}. We apply the finite difference method and define both spatial and temporal splitting 
\begin{equation}\label{grid_definition}
\Delta x=\frac{x_{\rm{max}}}{n_{x}-1},\, \Delta t=\frac{t_{\rm{max}}}{n_{t}-1},
\end{equation}
with $x_{\rm{max}}$ and $t_{\rm{max}}$ as limits for spatial and temporal dimension and $n_{x}$ and $n_{t}$ denoting the total number of spatial and temporal nodes of
the created grid. Using the definition of infinitesimal grid elements as (\ref{grid_definition}) we write the scheme for numerical solution to equation (\ref{prepis_surface_density_profile_equation}) as \citep{Bath}
\begin{equation}\label{final_scheme}
U_{i}^{m+1}=\frac{12 \nu \Delta t}{x_{i}^{2} \Delta x^{2}}(U_{i+1}^{m}+U_{i-1}^{m}-2U_{i}^{m})+U_{i}^{m},
\end{equation}
with truncation errors $o(\Delta x^{2})$ and $o(\Delta t)$ in spatial and temporal direction of the grid.
The stability of the solution given by (\ref{final_scheme}) must adhere to von-Neumann stability criterion which is in this case defined as \citep{Bath}
\begin{equation}\label{von_Neumann}
\frac{12 \nu \Delta t}{x_{i}^{2}\Delta x^{2}}<\frac{1}{2}.
\end{equation}
To get a physically realistic estimate of the value of kinematic viscosity playing crucial role in the stability of numerical solution to equation (\ref{surfce_density_profile_equation}) we use its definition \citep{frank,2019NewA...70....7M}
\begin{equation}\label{kinematicka_viskozita}
\nu=\alpha c_{\rm{s}} H,
\end{equation}
where $\alpha$ is the viscosity parameter introduced in the Shakura--Sunyaev standard accretion disc model, fulfilling the inequality $\alpha \lesssim 1$; $c_{\rm{s}}$ is speed of sound and $H(R)$ the typical scale height of the accretion disc. We presume the material circulating in the accretion disc to be that of an ideal gas and 
\begin{equation}
c_{\rm{s}}^{2}=\frac{k_{\rm{b}}T}{\mu m_{\rm{p}}}
\end{equation}
with $\mu$ the mean molecular weight and $m_{\rm{p}}$ the mass of proton. To express the typical scale height of the accretion disc as well as its temperature profile together with the accretion luminosity (to conclude the set of equations needed to obtain the estimate of the value of kinematic viscosity) we used standard definitions \citep{frank}. In the adopted approach the accretion luminosity is limited by the Eddington luminosity of the respective star that has been tidally disrupted. 

\begin{figure}[tbh!]
\centerline{\includegraphics[scale=0.49]{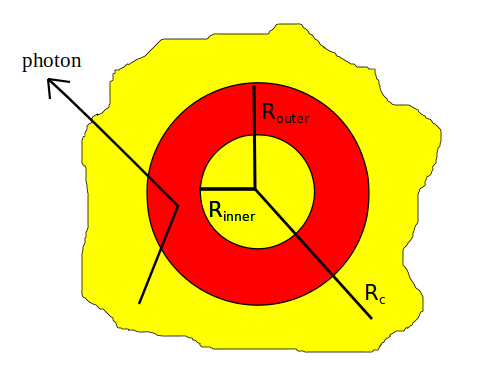}}
\caption{Sketch of the geometrical set-up (top view of the system). The red area depicts the gradually spreading accretion disc with its inner $R_{\rm{inner}}$ and outer $R_{\rm{outer}}$ radii. The yellow area represents the view of a hot corona (spherical radius $R_{\rm{c}}$) projected onto the equatorial plane that illuminates the underlying fluid ring a gives rise to the emerging spectral feature. The arrow indicates direction towards a distant observer.}\label{system_set-up}
\end{figure}

We note that we were looking for the estimate value of the kinematic viscosity independent on the radial component of the accretion disc, i.e. having constant value. 
Figure \ref{boundary_conditions} depicts the surface density profile for the different boundary conditions. The top panel shows the surface density profile to vanish once the inner radius of the accretion disc $R_{\rm{inner}}$ reaches (formally) the surface $R=R_{\rm{g}}$; the bottom panel assumes the value of the surface density profile to drop to zero at the innermost stable circular orbit (ISCO), which in case of a Schwarzschild-type SMBH equals to $6R_{\rm{g}}$. The latter case allows us to achieve better resolution considering the fact it stabilizes the solution of eqs.\ (\ref{final_scheme})--(\ref{von_Neumann}).

\section{Reflection spectral line from evolving accretion rings}\label{section4}
Within the context of a toy-model we consider the radiation in the spectral line to be produced by reflection on the accretion ring immersed in a corona (see Figure \ref{system_set-up}). We focus on modeling of the observed spectral line time evolution. We employ an approximation to account the relativistic effects on the energy and the direction of propagation of the photons.

We define the redshift factor \citep[e.g.][]{2006AN....327..961K}
\begin{equation}\label{redshift_factor}
g=\frac{\nu_{\rm{observed}}}{\nu_{\rm{local}}},
\end{equation}
with $\nu_{\rm{observed}}$ and $\nu_{\rm{local}}$ being the radiation frequency measured by the observer in infinity and in the close proximity of the accretion disc respectively. Aiming to take the relativistic effects into account in the first approximation we adopt the redshift factor \citep{pechacek}
\begin{equation}\label{g_gtr}
g=\\
\frac{\sqrt{R(R-3)}}{R+\sin(\varphi)\sin(I)\sqrt{R-2+4(1+\cos(\varphi)\sin(I))^{-1}}},
\end{equation}
where $R$ stands for the radial coordinate, $\varphi$ for the azimuthal coordinate and $I$ for the inclination paramater of the system.

Assuming the accretion disc or a ring of matter radiating at a single frequency, $\nu_{0}=\nu_{\rm{local}}$, we can represent the intrinsic line by the Dirac $\delta$-function, 
\begin{equation}\label{local_intesity_prescription}
I_{\nu}\approx\delta(\nu-\nu_{0})\frac{1}{R^{p}}, \, \delta(\nu-\nu_{0}) \approx \lim_{a \rightarrow 0}\exp\bigg[-\bigg(\frac{\nu-\nu_{0}}{a}\bigg)^{2}\bigg], 
\end{equation}
where $p$ is a parameter depending on the value of the outer radius of the corona surrounding the accretion disc, taking values $p = 0$ for $R < R_{\rm{c}}$ , $p = 2$ for $R \approx R_{\rm{c}}$ and $p = 3$ for $R > R_{\rm{c}}$ \citep{fabian}. We approach the evaluation of the total observed radiation flux as a superposition of the flux contributions from infinitesimally narrow rings. To this end we follow the formula 
\begin{equation}
F_{\nu}\approx \int I_{\nu}\,dS,
\end{equation}
with the surface element is positioned on the surface of the accretion disc, $dS = R\,dR\,d\varphi$, and ${I_{\nu}}/{\nu^{3}}={\rm constant}$ together with eqs.\ (\ref{g_gtr})--(\ref{local_intesity_prescription}).

\begin{figure*}[tbh!]
\centering
\hfill~\includegraphics[width=0.32\textwidth]{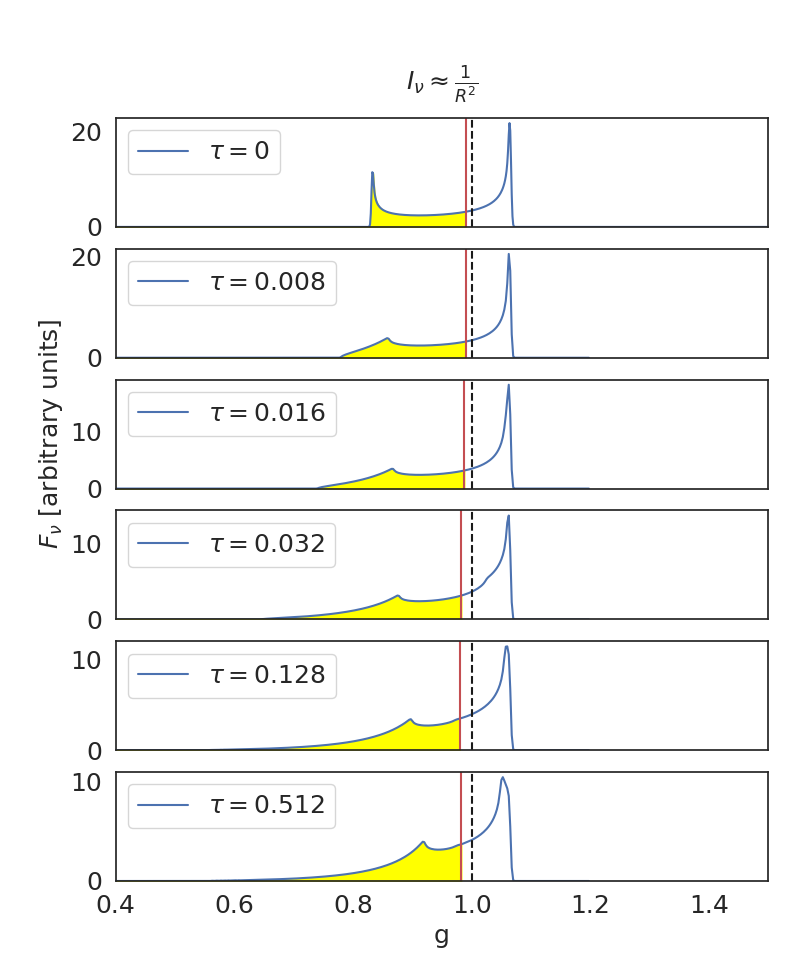}~\includegraphics[width=0.32\textwidth]{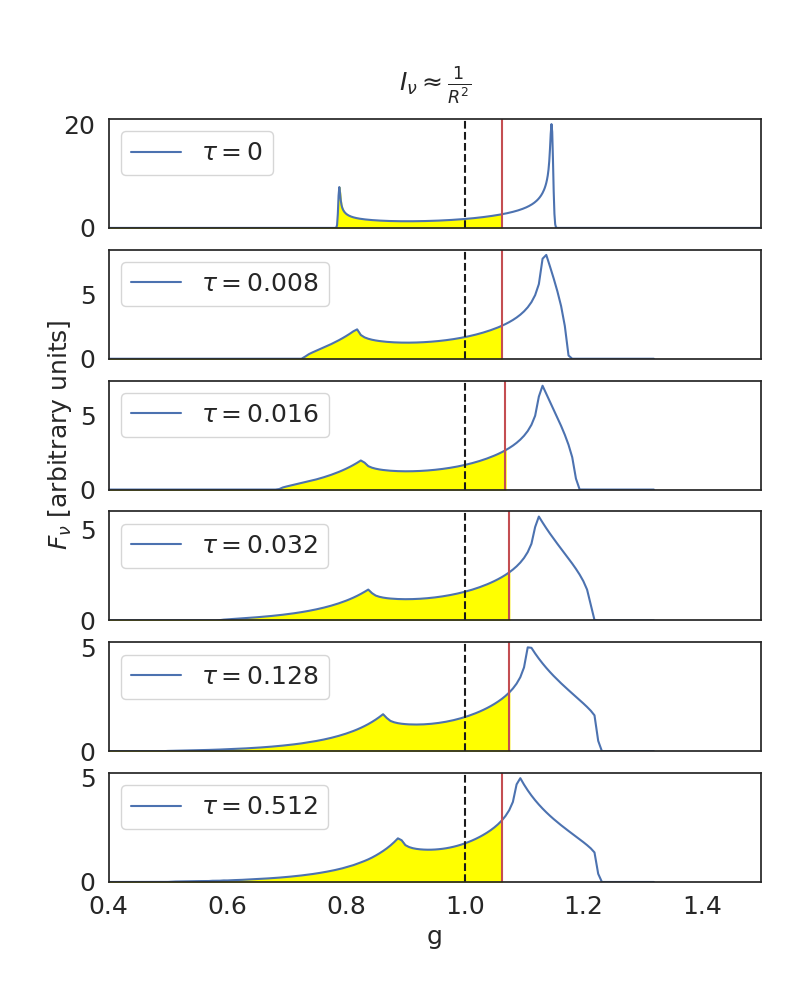}~\includegraphics[width=0.32\textwidth]{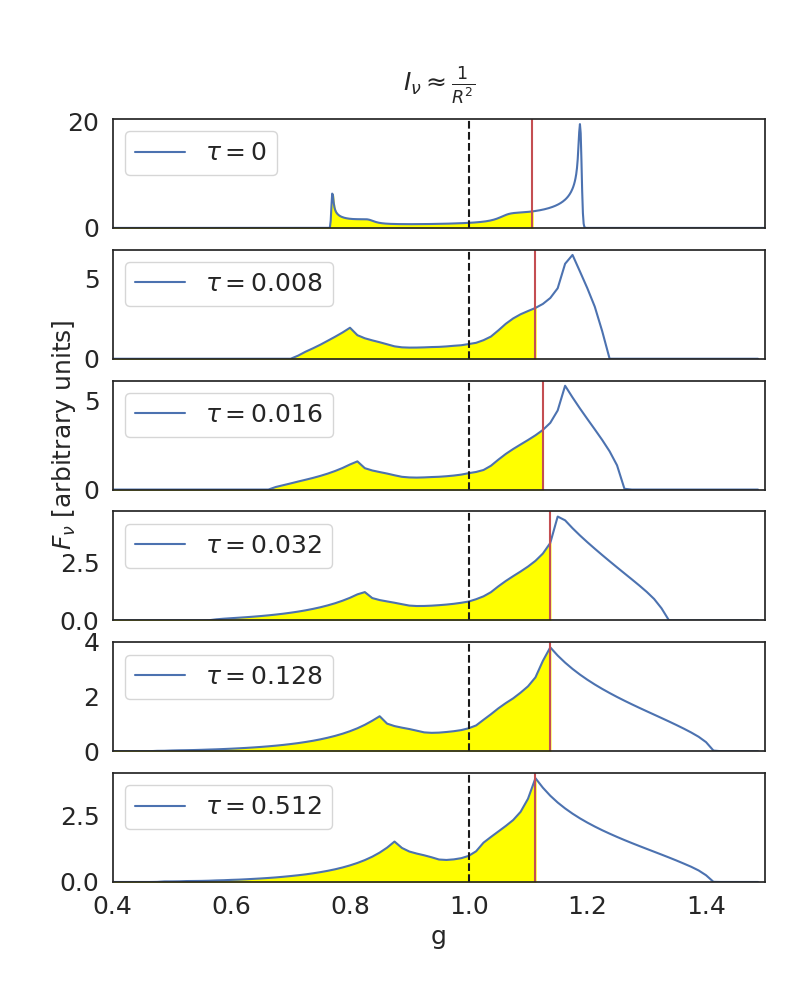}\hfill~\\
\hfill~\includegraphics[width=0.32\textwidth]{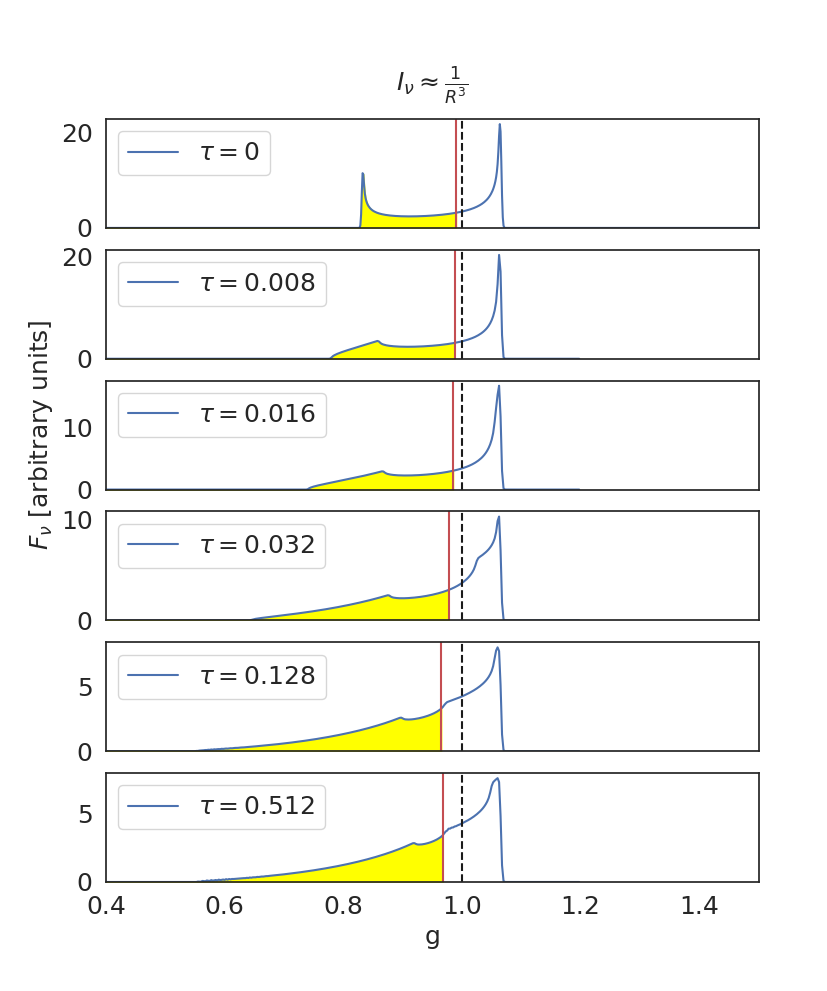}~\includegraphics[width=0.32\textwidth]{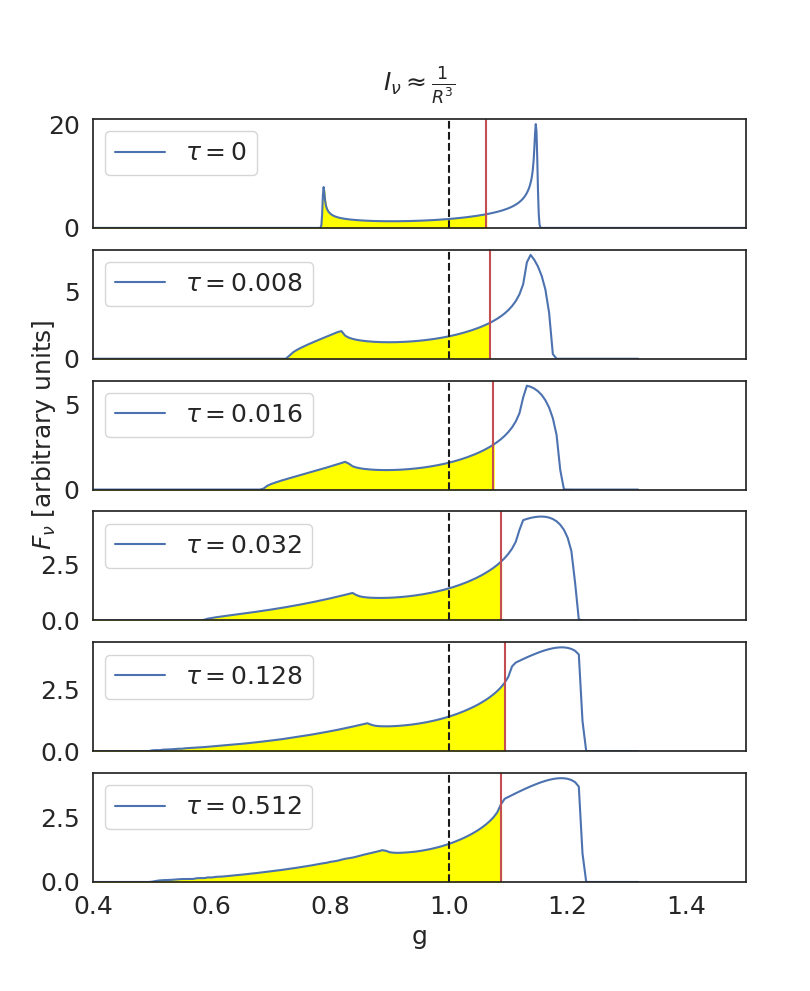}~\includegraphics[width=0.32\textwidth]{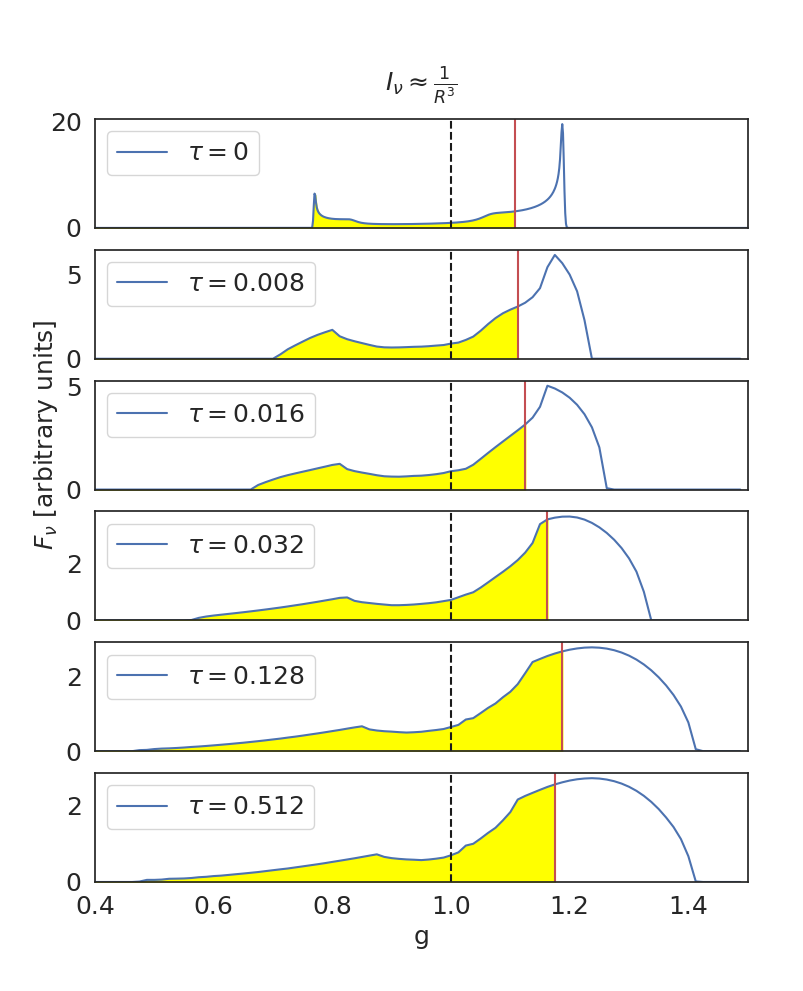}\hfill~
\caption{The modelled evolution of the relativistic spectral line profile as predicted for a viscously spreading accretion ring. The view angle is $35^\circ$ (left column),  $60^\circ$ (middle column), and $85^\circ$ (right column). The intrinsic radial emissivity profile is $I_{\nu}(R)\propto1/R^2$ (top row) represents the case of radially less compact illumination, while the more steep dependence $\propto1/R^3$ (bottom row) illustrates the effect of a compact corona. See the text for further details and the values of other parameters.}
\label{fig3}
\end{figure*}

In order to focus on the relativistic effects in the formation of the observed spectral line profiles reflected by the evolving accretion ring, we have limited our choice of the model parameters in such a way that the tidal radius remains close to the horizon,
\begin{equation}\label{threshold}
R_{\rm{tidal}} \lesssim 50 R_{\rm{g}}.
\end{equation}

The surface density profile of the initial mass ring at the tidal radius, $R_{0} = R_{\rm{tidal}}$, is evolved according to eq.\ (\ref{surfce_density_profile_equation}). Therefore, the outer radius of the accretion disc comes out more than $\sim2.5$ times that of the position of the initial mass ring. We thus set the threshold value for the value of the tidal radius (\ref{threshold}) as we aim to capture the spectral line profile time evolution after the occurrence of the TDE in the close proximity of the central object in the order of a few tens of gravitational radii; the relativistic effects are negligible farther away from SMBH. 

In the example shown here we assume the disrupted star to be of a solar type ($M_{\ast}= 1M_{\odot}$ and $R_{\ast}= 1R_{\odot}$); we then vary the mass of the SMBH over the representative values, including $M_{\rm{SMBH}}=4\times10^6M_{\odot}$. To estimate the size of the accretion discs we employed the boundary condition $\Sigma(R_{\rm{ISCO}}=6 R_{\rm{g}},t)= \Sigma (R_{\rm{outer}},t) = 0$, which ensures the surface density profile to vanish at ISCO. Figure \ref{fig3} depicts the expected spectral line temporal evolution for different inclination angles of the source and different prescription for the intrinsic (local) intensity, as described by eq.\ (\ref{local_intesity_prescription}). 

As mentioned above, the approximation scheme \citep{pechacek} is employed to account for the relativistic effects on light rays propagating from the spreading ring or an accretion disc towards a distant observer. We calculate the position of the broad line centroid energy, as it also changes with time during the accretion ring evolution (marked by the red line in Fig.\ \ref{fig3} panels; we also indicate the intrinsic (local frame) energy where the line has been emitted by the black dotted line). The reflection spectral features on the remnant trail resembles the situation studied previously in the context of AGN accretion rings \citep{2011MNRAS.418..276S,2017ApJS..229...40P}.

We observe that the case of only small inclination, $I=35\deg$ (almost pole-on view), exhibits more than half of the line radiation energy redshifted during the entire evolution cycle (see the part of the predicted profile indicated by the yellow colour; compare the left column vs.\ the right one in the plots). The corresponding plots for moderate ($I=60\deg$) and high inclination ($I=85\deg$, i.e.\ almost edge-on view) show more than half of the profile to be blue-shifted during the entire interval. Time steps are measured in dimensionless geometrized units ($\tau \in$ [0, 0.512]), thus corresponding proportionally to SMBH mass in physical units.

\begin{figure}[tbh!]
\centerline{\includegraphics[width=0.49\textwidth]{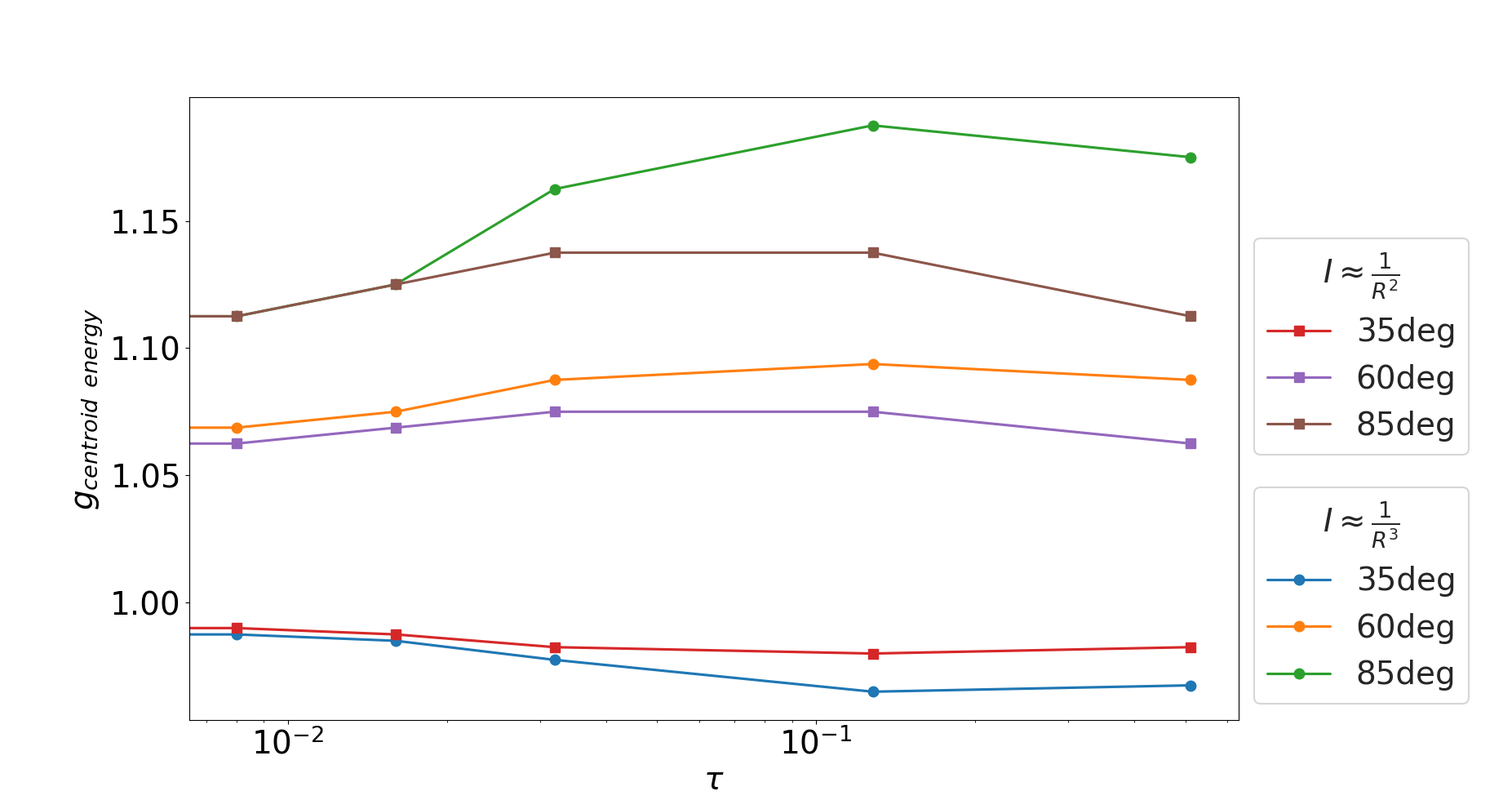}}
\caption{Centroid energy of the spectral line reflected from the equatorial gaseous accretion ring, as it spreads with time towards both higher and lower radii. Six cases are shown with the model parameters identical to the previous discussion in Fig.\ \ref{fig3} and labelled on the right side of the plot. The increase or the decrease of the centroid energy occur with time, depending on the mutual interplay of the model parameters, especially the inclination angle of the observer and the central compactness of the primary illumination of the accretion ring (see the text for further description).}
\label{fig4}
\end{figure}

Finally, Figure \ref{fig4} summarizes the dependency with time of the evolving energy centroid from the previous Fig.\ \ref{fig3} (indicated by the red line). While mapping the detailed profile of the spectral feature is probably beyond the expected resolution of near-future detections, the centroid energy reflects the overall position of the line and its change appears to be indicative of the parameters with a promising sensitivity. However, a certain degeneracy of the parameter values is apparent, namely, between the inclination $I$ (characterizing the observer's view angle), the radial index $p$ of the emissivity dependence (characterizing the compactness of the primary illumination by the corona), and the prescription for the inner boundary torque \citep[see][for further examples]{stolc2019}.

\section{Conclusions}\label{section6} 
We discussed a simple approximation scenario where a relativistically broadened and skewed spectral feature evolves in time due to viscous spreading of the underlying accretion ring. We suggested that this model may in near future form a basis to explore properties of such spectral features from Tidal Disruption Events occurring in AGN, where the remnant material from a disrupted star becomes embedded into a preexisting corona and illuminated by the primary X-rays. Especially for high inclination angles and late stages of the ring evolution we notice the development of an extended blue wing of the predicted line; the emerging feature adopts the shape quite different from the simple double-horn line expected from accretion discs with a fixed location of the edges. 

Our scheme has been meant as a qualitative treatment which captures the basic effects (energy shifts and the viscous spreading of the disc), however, it cannot take quantitative details into consideration and it also neglects the role of frame dragging of a rotating black hole. The latter complexities are deferred to the future work. In this study we adopted the most simplistic approach: we considered a fully circularized trail of the remnant material, evolving just by action of $\alpha$-viscosity in the Newtonian regime, and we included the relativistic effects by an approximation scheme. One can immediately identify a number of important aspects that have been neglected in our tentative considerations: (i)~the initial configuration of the accretion ring should consider its eccentricity and then follow the gradual circularization rather than assuming zero eccentricity {\it ab initio}; (ii)~the presumed hot corona should contribute to the line-profile smearing by Comptonization, which will require to perform a more sophisticated radiation transfer of the reflection spectrum; (iii)~effects of General Relativity (especially the role of frame-dragging) should be calculated in a self-consistent manner by adopting the relativistic formalism of Kerr solution; (iv)~the influence of magnetic fields should be also included, so that the viscosity prescription reflects the dominant effect of magneto-rotational instability; (v)~the spectral line profile should be combined with the underlying continuum that has to be also modelled and its parameters constrained by a simultaneous fitting procedure together with the line(s). All these factors need to be taken into account in order to produce an astrophysically realistic scenario, however, despite the above-mentioned neglections, the adopted scheme allowed us to demonstrate some of the expected effects on the emerging spectral features in a simplified approach.

As mentioned in the Sect.\ \ref{section4} the study of spectral line profile time evolution suggests a way how to infer the inclination of the radiation source as well as the method to constrain the size of the accretion disc, namely, the distance of the circulating debris from the SMBH and its evolving radial extent. We have postponed the prospects of a more complete analysis involving the SMBH spin and charge. It will then be necessary to use the redshift factor corresponding to the rotating (Kerr) metric, or even the rotating and electrically charged (Kerr--Newman) metric \citep{2006AN....327..961K,2019arXiv190106507K}. The initial moments after the trail formation will require to consider an eccentric ring. Furthermore, it will be necessary to adapt the equations (\ref{continuity_equation}) and (\ref{momentum_conservation_equation}) leading to the equation for the surface density profile of the accretion disc (\ref{surfce_density_profile_equation}) to a more complicated framework of General Relativity. 

\subsection*{Author contributions}

M.\ \v{S}tolc performed calculations, prepared graphs and wrote the main part of the text. V.\ Karas provided additional analysis and contributed to the text of the article in all its sections.

\subsection*{Financial disclosure}
\fundingAgency{Czech Science Foundation -- Deutsche Forschungsgemeinschaft} No.~\fundingNumber{19-01137J}, and the COST Action MP1304 ``Exploring Fundamental Physics with Compact Stars'' grant of the \fundingAgency{Czech Ministry of Education, Youth and Sports} No.~\fundingNumber{LD15061}.

\subsection*{Conflict of interest}

The authors declare no potential conflict of interests.

\section*{Acknowledgments}
We thank an anonymous referee for helpful suggestions. We acknowledge the collaboration grant of the Czech Science Foundation -- Deutsche Forschungsgemeinschaft (No.~19-01137J), and the COST Action MP1304 ``Exploring Fundamental Physics with Compact Stars'' of the Czech Ministry of Education, Youth and Sports (No.~LD15061). 

\section*{Author Biography}
During the preparation of this contribution \textbf{Marcel \v{S}tolc} was a student working towards Master Degree at the Astronomical Institute of the Charles University in Prague (Faculty of Mathematics and Physics). Currently he is Ph.D. student in \href{http://astro.cas.cz}{Prague Relativistic Astrophysics} working group. \textbf{Vladim\'{\i}r Karas} is senior researcher and a supervisor working in the Astronomical Institute of the Czech Academy of Sciences.

\bibliography{stolc-karas-IBWS-2019}

\begin{thebibliography}{}

\bibitem [\protect \citeauthoryear {%
{Bath}%
\ \BBA {} {Pringle}%
}{%
{Bath}%
\ \BBA {} {Pringle}%
}{%
{\protect \APACyear {1981}}%
}]{%
Bath}
\APACinsertmetastar {%
Bath}%
\begin{APACrefauthors}%
{Bath}, G\BPBI T.%
\BCBT {}\ \BBA {} {Pringle}, J\BPBI E.%
\end{APACrefauthors}%
\unskip\
\newblock
\APACrefYearMonthDay{1981}{}{},
\newblock
\unskip
\newblock
\APACjournalVolNumPages{MNRAS}{194}{}{967-986}.
\newblock
\begin{APACrefURL} \url{http://adsabs.harvard.edu/abs/1981MNRAS.194..967B}
  \end{APACrefURL}
\PrintBackRefs{\CurrentBib}

\bibitem [\protect \citeauthoryear {%
{Chang}%
\ \BBA {} {Choi}%
}{%
{Chang}%
\ \BBA {} {Choi}%
}{%
{\protect \APACyear {2002}}%
}]{%
2002JKAS...35..123C}
\APACinsertmetastar {%
2002JKAS...35..123C}%
\begin{APACrefauthors}%
{Chang}, H\BHBI Y.%
\BCBT {}\ \BBA {} {Choi}, C\BHBI S.%
\end{APACrefauthors}%
\unskip\
\newblock
\APACrefYearMonthDay{2002}{}{},
\newblock
\unskip
\newblock
\APACjournalVolNumPages{Journal of Korean Astronomical Society}{35}{}{123-130}.
\newblock
\begin{APACrefURL} \url{https://ui.adsabs.harvard.edu/abs/2002JKAS...35..123C}
  \end{APACrefURL}
\PrintBackRefs{\CurrentBib}

\bibitem [\protect \citeauthoryear {%
{Coughlin}%
\ \BBA {} {Nixon}%
}{%
{Coughlin}%
\ \BBA {} {Nixon}%
}{%
{\protect \APACyear {2019}}%
}]{%
2019arXiv190703034C}
\APACinsertmetastar {%
2019arXiv190703034C}%
\begin{APACrefauthors}%
{Coughlin}, E\BPBI R.%
\BCBT {}\ \BBA {} {Nixon}, C\BPBI J.%
\end{APACrefauthors}%
\unskip\
\newblock
\APACrefYearMonthDay{2019}{Jul}{},
\newblock
\unskip
\newblock
\APACjournalVolNumPages{ApJL, {\rm submitted}}{}{}{arXiv:1907.03034}.
\PrintBackRefs{\CurrentBib}

\bibitem [\protect \citeauthoryear {%
{Cummings}%
\ \protect \BOthers {.}}{%
{Cummings}%
\ \protect \BOthers {.}}{%
{\protect \APACyear {2011}}%
}]{%
cummings}
\APACinsertmetastar {%
cummings}%
\begin{APACrefauthors}%
{Cummings}, J\BPBI R.%
, {Barthelmy}, S\BPBI D.%
, {Beardmore}, A\BPBI P.%
\ et al.\end{APACrefauthors}%
\unskip\
\newblock
\APACrefYearMonthDay{2011}{}{},
\newblock
\unskip
\newblock
\APACjournalVolNumPages{GRB Coordinates Network, Circular Service, No.~11823,
  \#1 (2011)}{11823}{}{}.
\newblock
\begin{APACrefURL} \url{http://adsabs.harvard.edu/abs/2011GCN.11823....1C}
  \end{APACrefURL}
\PrintBackRefs{\CurrentBib}

\bibitem [\protect \citeauthoryear {%
{Dai}%
, {McKinney}%
, {Roth}%
, {Ramirez-Ruiz}%
\BCBL {}\ \BBA {} {Miller}%
}{%
{Dai}%
\ \protect \BOthers {.}}{%
{\protect \APACyear {2018}}%
}]{%
2018ApJ...859L..20D}
\APACinsertmetastar {%
2018ApJ...859L..20D}%
\begin{APACrefauthors}%
{Dai}, L.%
, {McKinney}, J\BPBI C.%
, {Roth}, N.%
, {Ramirez-Ruiz}, E.%
\BCBL {}\ \BBA {} {Miller}, M\BPBI C.%
\end{APACrefauthors}%
\unskip\
\newblock
\APACrefYearMonthDay{2018}{}{},
\newblock
\unskip
\newblock
\APACjournalVolNumPages{ApJ}{859}{2}{L20}.
\newblock
\begin{APACrefURL} \url{https://ui.adsabs.harvard.edu/abs/2018ApJ...859L..20D}
  \end{APACrefURL}
\PrintBackRefs{\CurrentBib}

\bibitem [\protect \citeauthoryear {%
{Evans}%
\ \BBA {} {Kochanek}%
}{%
{Evans}%
\ \BBA {} {Kochanek}%
}{%
{\protect \APACyear {1989}}%
}]{%
1989ApJ...346L..13E}
\APACinsertmetastar {%
1989ApJ...346L..13E}%
\begin{APACrefauthors}%
{Evans}, C\BPBI R.%
\BCBT {}\ \BBA {} {Kochanek}, C\BPBI S.%
\end{APACrefauthors}%
\unskip\
\newblock
\APACrefYearMonthDay{1989}{{\APACmonth{11}}}{},
\newblock
\unskip
\newblock
\APACjournalVolNumPages{\apjl}{346}{}{L13-L16}.
\newblock
\begin{APACrefDOI} \doi{10.1086/185567} \end{APACrefDOI}
\PrintBackRefs{\CurrentBib}

\bibitem [\protect \citeauthoryear {%
{Fabian}%
, {Rees}%
, {Stella}%
\BCBL {}\ \BBA {} {White}%
}{%
{Fabian}%
\ \protect \BOthers {.}}{%
{\protect \APACyear {1989}}%
}]{%
fabian}
\APACinsertmetastar {%
fabian}%
\begin{APACrefauthors}%
{Fabian}, A\BPBI C.%
, {Rees}, M\BPBI J.%
, {Stella}, L.%
\BCBL {}\ \BBA {} {White}, N\BPBI E.%
\end{APACrefauthors}%
\unskip\
\newblock
\APACrefYearMonthDay{1989}{}{},
\newblock
\unskip
\newblock
\APACjournalVolNumPages{MNRAS}{238}{}{729-736}.
\newblock
\begin{APACrefURL} \url{https://ui.adsabs.harvard.edu/abs/1989MNRAS.238..729F}
  \end{APACrefURL}
\PrintBackRefs{\CurrentBib}

\bibitem [\protect \citeauthoryear {%
{Frank}%
, {King}%
\BCBL {}\ \BBA {} {Raine}%
}{%
{Frank}%
\ \protect \BOthers {.}}{%
{\protect \APACyear {2002}}%
}]{%
frank}
\APACinsertmetastar {%
frank}%
\begin{APACrefauthors}%
{Frank}, J.%
, {King}, A.%
\BCBL {}\ \BBA {} {Raine}, D.%
\end{APACrefauthors}%
\unskip\
\newblock
\APACrefYearMonthDay{2002}{}{},
\newblock
\unskip
\newblock
\APACjournalVolNumPages{(Cambridge University Press; Cambridge)}{}{}{}.
\PrintBackRefs{\CurrentBib}

\bibitem [\protect \citeauthoryear {%
{Gallegos-Garcia}%
, {Law-Smith}%
\BCBL {}\ \BBA {} {Ramirez-Ruiz}%
}{%
{Gallegos-Garcia}%
\ \protect \BOthers {.}}{%
{\protect \APACyear {2018}}%
}]{%
2018ApJ...857..109G}
\APACinsertmetastar {%
2018ApJ...857..109G}%
\begin{APACrefauthors}%
{Gallegos-Garcia}, M.%
, {Law-Smith}, J.%
\BCBL {}\ \BBA {} {Ramirez-Ruiz}, E.%
\end{APACrefauthors}%
\unskip\
\newblock
\APACrefYearMonthDay{2018}{Apr}{},
\newblock
\unskip
\newblock
\APACjournalVolNumPages{ApJ}{857}{2}{109}.
\PrintBackRefs{\CurrentBib}

\bibitem [\protect \citeauthoryear {%
{Hills}%
}{%
{Hills}%
}{%
{\protect \APACyear {1975}}%
}]{%
hills}
\APACinsertmetastar {%
hills}%
\begin{APACrefauthors}%
{Hills}, J\BPBI G.%
\end{APACrefauthors}%
\unskip\
\newblock
\APACrefYearMonthDay{1975}{}{},
\newblock
\unskip
\newblock
\APACjournalVolNumPages{Nature}{254}{}{295-298}.
\newblock
\begin{APACrefURL} \url{http://adsabs.harvard.edu/abs/1975Natur.254..295H}
  \end{APACrefURL}
\PrintBackRefs{\CurrentBib}

\bibitem [\protect \citeauthoryear {%
{Kara}%
, {Miller}%
, {Reynolds}%
\BCBL {}\ \BBA {} {Dai}%
}{%
{Kara}%
\ \protect \BOthers {.}}{%
{\protect \APACyear {2016}}%
}]{%
kara}
\APACinsertmetastar {%
kara}%
\begin{APACrefauthors}%
{Kara}, E.%
, {Miller}, J\BPBI M.%
, {Reynolds}, C.%
\BCBL {}\ \BBA {} {Dai}, L.%
\end{APACrefauthors}%
\unskip\
\newblock
\APACrefYearMonthDay{2016}{}{},
\newblock
\unskip
\newblock
\APACjournalVolNumPages{Nature}{535}{}{388-390}.
\newblock
\begin{APACrefURL} \url{http://adsabs.harvard.edu/abs/2016Natur.535..388K}
  \end{APACrefURL}
\PrintBackRefs{\CurrentBib}

\bibitem [\protect \citeauthoryear {%
{Karas}%
}{%
{Karas}%
}{%
{\protect \APACyear {2006}}%
}]{%
2006AN....327..961K}
\APACinsertmetastar {%
2006AN....327..961K}%
\begin{APACrefauthors}%
{Karas}, V.%
\end{APACrefauthors}%
\unskip\
\newblock
\APACrefYearMonthDay{2006}{}{},
\newblock
\unskip
\newblock
\APACjournalVolNumPages{AN}{327}{10}{961-968}.
\newblock
\begin{APACrefURL} \url{https://ui.adsabs.harvard.edu/abs/2006AN....327..961K}
  \end{APACrefURL}
\PrintBackRefs{\CurrentBib}

\bibitem [\protect \citeauthoryear {%
{Karas}%
, {Dov{\v c}iak}%
, {Kunneriath}%
, {Yu}%
\BCBL {}\ \BBA {} {Zhang}%
}{%
{Karas}%
\ \protect \BOthers {.}}{%
{\protect \APACyear {2014}}%
}]{%
2014bhns.work..129K}
\APACinsertmetastar {%
2014bhns.work..129K}%
\begin{APACrefauthors}%
{Karas}, V.%
, {Dov{\v c}iak}, M.%
, {Kunneriath}, D.%
, {Yu}, W.%
\BCBL {}\ \BBA {} {Zhang}, W.%
\end{APACrefauthors}%
\unskip\
\newblock
\APACrefYearMonthDay{2014}{}{},
\newblock
{\BBOQ}\APACrefatitle {{Tidal disruption events from a nuclear star cluster as
  possible origin of transient relativistic spectral lines near SMBH}} {{Tidal
  disruption events from a nuclear star cluster as possible origin of transient
  relativistic spectral lines near SMBH}}.{\BBCQ}
\newblock
\BIn{} \APACrefbtitle {{\it Proceedings of RAGtime: Workshops on black holes
  and neutron stars}; Z.~Stuchl{\'{\i}}k, G.~T{\"o}r{\"o}k and T.~Pech{\'a}{\v
  c}ek, editors (Silesian University, Opava),} {{\it Proceedings of RAGtime:
  Workshops on black holes and neutron stars}; Z.~Stuchl{\'{\i}}k,
  G.~T{\"o}r{\"o}k and T.~Pech{\'a}{\v c}ek, editors (Silesian University,
  Opava),}\ \BPG~129-136.
\newblock
\begin{APACrefURL} \url{https://ui.adsabs.harvard.edu/abs/2014bhns.work..129K}
  \end{APACrefURL}
\PrintBackRefs{\CurrentBib}

\bibitem [\protect \citeauthoryear {%
{Karas}%
, {Svoboda}%
\BCBL {}\ \BBA {} {Zaja\v{c}ek}%
}{%
{Karas}%
\ \protect \BOthers {.}}{%
{\protect \APACyear {2019}}%
}]{%
2019arXiv190106507K}
\APACinsertmetastar {%
2019arXiv190106507K}%
\begin{APACrefauthors}%
{Karas}, V.%
, {Svoboda}, J.%
\BCBL {}\ \BBA {} {Zaja\v{c}ek}, M.%
\end{APACrefauthors}%
\unskip\
\newblock
\APACrefYearMonthDay{2019}{}{},
\newblock
\unskip
\newblock
\APACjournalVolNumPages{Lecture Notes (Summer School, Univ. of
  Bremen)}{}{}{arXiv:1901.06507}.
\newblock
\begin{APACrefURL} \url{https://ui.adsabs.harvard.edu/abs/2019arXiv190106507K}
  \end{APACrefURL}
\PrintBackRefs{\CurrentBib}

\bibitem [\protect \citeauthoryear {%
{Kato}%
, {Fukue}%
\BCBL {}\ \BBA {} {Mineshige}%
}{%
{Kato}%
\ \protect \BOthers {.}}{%
{\protect \APACyear {2008}}%
}]{%
2008bhad.book.....K}
\APACinsertmetastar {%
2008bhad.book.....K}%
\begin{APACrefauthors}%
{Kato}, S.%
, {Fukue}, J.%
\BCBL {}\ \BBA {} {Mineshige}, S.%
\end{APACrefauthors}%
\unskip\
\newblock
\APACrefYear{2008},
\newblock
\APACrefbtitle {{Black-Hole Accretion Disks -- Towards a New Paradigm (Kyoto
  University Press, Kyoto, Japan)}} {{Black-Hole Accretion Disks -- Towards a
  New Paradigm (Kyoto University Press, Kyoto, Japan)}}.
\newblock
\begin{APACrefURL} \url{https://ui.adsabs.harvard.edu/abs/2008bhad.book.....K}
  \end{APACrefURL}
\PrintBackRefs{\CurrentBib}

\bibitem [\protect \citeauthoryear {%
{Komossa}%
}{%
{Komossa}%
}{%
{\protect \APACyear {2015}}%
}]{%
Komossa}
\APACinsertmetastar {%
Komossa}%
\begin{APACrefauthors}%
{Komossa}, S.%
\end{APACrefauthors}%
\unskip\
\newblock
\APACrefYearMonthDay{2015}{}{},
\newblock
\unskip
\newblock
\APACjournalVolNumPages{Journal of High Energy Astrophysics}{7}{}{148-157}.
\newblock
\begin{APACrefURL} \url{http://adsabs.harvard.edu/abs/2015JHEAp...7..148K}
  \end{APACrefURL}
\PrintBackRefs{\CurrentBib}

\bibitem [\protect \citeauthoryear {%
{Komossa}%
}{%
{Komossa}%
}{%
{\protect \APACyear {2017}}%
}]{%
2017AN....338..256K}
\APACinsertmetastar {%
2017AN....338..256K}%
\begin{APACrefauthors}%
{Komossa}, S.%
\end{APACrefauthors}%
\unskip\
\newblock
\APACrefYearMonthDay{2017}{}{},
\newblock
\unskip
\newblock
\APACjournalVolNumPages{Astronomische Nachrichten}{338}{256}{256-261}.
\newblock
\begin{APACrefURL} \url{https://ui.adsabs.harvard.edu/abs/2017AN....338..256K}
  \end{APACrefURL}
\PrintBackRefs{\CurrentBib}

\bibitem [\protect \citeauthoryear {%
{Komossa}%
\ \protect \BOthers {.}}{%
{Komossa}%
\ \protect \BOthers {.}}{%
{\protect \APACyear {2004}}%
}]{%
2004ApJ...603L..17K}
\APACinsertmetastar {%
2004ApJ...603L..17K}%
\begin{APACrefauthors}%
{Komossa}, S.%
, {Halpern}, J.%
, {Schartel}, N.%
, {Hasinger}, G.%
, {Santos-Lleo}, M.%
\BCBL {}\ \BBA {} {Predehl}, P.%
\end{APACrefauthors}%
\unskip\
\newblock
\APACrefYearMonthDay{2004}{}{},
\newblock
\unskip
\newblock
\APACjournalVolNumPages{ApJL}{603}{}{L17-L20}.
\newblock
\begin{APACrefURL} \url{https://ui.adsabs.harvard.edu/abs/2004ApJ...603L..17K}
  \end{APACrefURL}
\PrintBackRefs{\CurrentBib}

\bibitem [\protect \citeauthoryear {%
{Luminet}%
\ \BBA {} {Marck}%
}{%
{Luminet}%
\ \BBA {} {Marck}%
}{%
{\protect \APACyear {1985}}%
}]{%
1985MNRAS.212...57L}
\APACinsertmetastar {%
1985MNRAS.212...57L}%
\begin{APACrefauthors}%
{Luminet}, J\BHBI P.%
\BCBT {}\ \BBA {} {Marck}, J\BHBI A.%
\end{APACrefauthors}%
\unskip\
\newblock
\APACrefYearMonthDay{1985}{}{},
\newblock
\unskip
\newblock
\APACjournalVolNumPages{MNRAS}{212}{}{57-75}.
\newblock
\begin{APACrefURL} \url{https://ui.adsabs.harvard.edu/abs/1985MNRAS.212...57L}
  \end{APACrefURL}
\PrintBackRefs{\CurrentBib}

\bibitem [\protect \citeauthoryear {%
{Magorrian}%
\ \protect \BOthers {.}}{%
{Magorrian}%
\ \protect \BOthers {.}}{%
{\protect \APACyear {1998}}%
}]{%
magorian}
\APACinsertmetastar {%
magorian}%
\begin{APACrefauthors}%
{Magorrian}, J.%
, {Tremaine}, S.%
, {Richstone}, D.%
\ et al.\end{APACrefauthors}%
\unskip\
\newblock
\APACrefYearMonthDay{1998}{}{},
\newblock
\unskip
\newblock
\APACjournalVolNumPages{AJ}{115}{}{2285-2305}.
\newblock
\begin{APACrefURL} \url{http://adsabs.harvard.edu/abs/1998AJ....115.2285M}
  \end{APACrefURL}
\PrintBackRefs{\CurrentBib}

\bibitem [\protect \citeauthoryear {%
{Mannheim}%
, {Grupe}%
, {Beuermann}%
, {Thomas}%
\BCBL {}\ \BBA {} {Fink}%
}{%
{Mannheim}%
\ \protect \BOthers {.}}{%
{\protect \APACyear {1996}}%
}]{%
1996rftu.proc..471M}
\APACinsertmetastar {%
1996rftu.proc..471M}%
\begin{APACrefauthors}%
{Mannheim}, K.%
, {Grupe}, D.%
, {Beuermann}, K.%
, {Thomas}, H\BPBI C.%
\BCBL {}\ \BBA {} {Fink}, H\BPBI H.%
\end{APACrefauthors}%
\unskip\
\newblock
\APACrefYearMonthDay{1996}{}{},
\newblock
{\BBOQ}\APACrefatitle {{Ultrasoft transient X-ray emission from AGN.}}
  {{Ultrasoft transient X-ray emission from AGN.}}{\BBCQ}
\newblock
\BIn{} H\BPBI U.~{Zimmermann}, J.~{Tr{\"u}mper}\BCBL {}\ \BBA {} H.~{Yorke}\
  (\BEDS), \APACrefbtitle {Roentgenstrahlung from the Universe,}
  {Roentgenstrahlung from the Universe,}\ \BPG~471-472.
\newblock
\begin{APACrefURL} \url{https://ui.adsabs.harvard.edu/abs/1996rftu.proc..471M}
  \end{APACrefURL}
\PrintBackRefs{\CurrentBib}

\bibitem [\protect \citeauthoryear {%
{Martin}%
, {Nixon}%
, {Pringle}%
\BCBL {}\ \BBA {} {Livio}%
}{%
{Martin}%
\ \protect \BOthers {.}}{%
{\protect \APACyear {2019}}%
}]{%
2019NewA...70....7M}
\APACinsertmetastar {%
2019NewA...70....7M}%
\begin{APACrefauthors}%
{Martin}, R\BPBI G.%
, {Nixon}, C\BPBI J.%
, {Pringle}, J\BPBI E.%
\BCBL {}\ \BBA {} {Livio}, M.%
\end{APACrefauthors}%
\unskip\
\newblock
\APACrefYearMonthDay{2019}{}{},
\newblock
\unskip
\newblock
\APACjournalVolNumPages{\na}{70}{}{7-11}.
\newblock
\begin{APACrefURL} \url{https://ui.adsabs.harvard.edu/abs/2019NewA...70....7M}
  \end{APACrefURL}
\PrintBackRefs{\CurrentBib}

\bibitem [\protect \citeauthoryear {%
{Montesinos Armijo}%
\ \BBA {} {de Freitas Pacheco}%
}{%
{Montesinos Armijo}%
\ \BBA {} {de Freitas Pacheco}%
}{%
{\protect \APACyear {2011}}%
}]{%
monte}
\APACinsertmetastar {%
monte}%
\begin{APACrefauthors}%
{Montesinos Armijo}, M.%
\BCBT {}\ \BBA {} {de Freitas Pacheco}, J\BPBI A.%
\end{APACrefauthors}%
\unskip\
\newblock
\APACrefYearMonthDay{2011}{}{},
\newblock
\unskip
\newblock
\APACjournalVolNumPages{ApJ}{736}{}{126}.
\newblock
\begin{APACrefURL} \url{http://adsabs.harvard.edu/abs/2011ApJ...736..126M}
  \end{APACrefURL}
\PrintBackRefs{\CurrentBib}

\bibitem [\protect \citeauthoryear {%
{Pech{\'a}{\v c}ek}%
, {Dov{\v c}iak}%
\BCBL {}\ \BBA {} {Karas}%
}{%
{Pech{\'a}{\v c}ek}%
\ \protect \BOthers {.}}{%
{\protect \APACyear {2005}}%
}]{%
pechacek}
\APACinsertmetastar {%
pechacek}%
\begin{APACrefauthors}%
{Pech{\'a}{\v c}ek}, T.%
, {Dov{\v c}iak}, M.%
\BCBL {}\ \BBA {} {Karas}, V.%
\end{APACrefauthors}%
\unskip\
\newblock
\APACrefYearMonthDay{2005}{}{},
\newblock
\unskip
\newblock
\APACjournalVolNumPages{Proceedings of RAGtime: Workshops on black holes and
  neutron stars (Silesian University, Opava)}{}{}{137-141}.
\newblock
\begin{APACrefURL} \url{http://adsabs.harvard.edu/abs/2005ragt.meet..137P}
  \end{APACrefURL}
\PrintBackRefs{\CurrentBib}

\bibitem [\protect \citeauthoryear {%
{Pugliese}%
\ \BBA {} {Stuchl{\'\i}k}%
}{%
{Pugliese}%
\ \BBA {} {Stuchl{\'\i}k}%
}{%
{\protect \APACyear {2017}}%
}]{%
2017ApJS..229...40P}
\APACinsertmetastar {%
2017ApJS..229...40P}%
\begin{APACrefauthors}%
{Pugliese}, D.%
\BCBT {}\ \BBA {} {Stuchl{\'\i}k}, Z.%
\end{APACrefauthors}%
\unskip\
\newblock
\APACrefYearMonthDay{2017}{Apr}{},
\newblock
\unskip
\newblock
\APACjournalVolNumPages{ApJS}{229}{2}{40}.
\newblock
\begin{APACrefURL} \url{https://ui.adsabs.harvard.edu/abs/2017ApJS..229...40P}
  \end{APACrefURL}
\PrintBackRefs{\CurrentBib}

\bibitem [\protect \citeauthoryear {%
{Rees}%
}{%
{Rees}%
}{%
{\protect \APACyear {1988}}%
}]{%
rees}
\APACinsertmetastar {%
rees}%
\begin{APACrefauthors}%
{Rees}, M\BPBI J.%
\end{APACrefauthors}%
\unskip\
\newblock
\APACrefYearMonthDay{1988}{}{},
\newblock
\unskip
\newblock
\APACjournalVolNumPages{Nature}{333}{}{523-528}.
\newblock
\begin{APACrefURL} \url{http://adsabs.harvard.edu/abs/1988Natur.333..523R}
  \end{APACrefURL}
\PrintBackRefs{\CurrentBib}

\bibitem [\protect \citeauthoryear {%
{Saxton}%
\ \protect \BOthers {.}}{%
{Saxton}%
\ \protect \BOthers {.}}{%
{\protect \APACyear {2012}}%
}]{%
2012A&A...541A.106S}
\APACinsertmetastar {%
2012A&A...541A.106S}%
\begin{APACrefauthors}%
{Saxton}, R\BPBI D.%
, {Read}, A\BPBI M.%
, {Esquej}, P.%
, {Komossa}, S.%
, {Dougherty}, S.%
, {Rodriguez-Pascual}, P.%
\BCBL {}\ \BBA {} {Barrado}, D.%
\end{APACrefauthors}%
\unskip\
\newblock
\APACrefYearMonthDay{2012}{}{},
\newblock
\unskip
\newblock
\APACjournalVolNumPages{A\&A}{541}{}{A106}.
\newblock
\begin{APACrefURL}
  \url{https://ui.adsabs.harvard.edu/abs/2012A%26A...541A.106S}
  \end{APACrefURL}
\PrintBackRefs{\CurrentBib}

\bibitem [\protect \citeauthoryear {%
{Sochora}%
, {Karas}%
, {Svoboda}%
\BCBL {}\ \BBA {} {Dov{\v{c}}iak}%
}{%
{Sochora}%
\ \protect \BOthers {.}}{%
{\protect \APACyear {2011}}%
}]{%
2011MNRAS.418..276S}
\APACinsertmetastar {%
2011MNRAS.418..276S}%
\begin{APACrefauthors}%
{Sochora}, V.%
, {Karas}, V.%
, {Svoboda}, J.%
\BCBL {}\ \BBA {} {Dov{\v{c}}iak}, M.%
\end{APACrefauthors}%
\unskip\
\newblock
\APACrefYearMonthDay{2011}{Nov}{},
\newblock
\unskip
\newblock
\APACjournalVolNumPages{MNRAS}{418}{1}{276-283}.
\newblock
\begin{APACrefURL} \url{https://ui.adsabs.harvard.edu/abs/2011MNRAS.418..276S}
  \end{APACrefURL}
\PrintBackRefs{\CurrentBib}

\bibitem [\protect \citeauthoryear {%
{\v{S}tolc}%
}{%
{\v{S}tolc}%
}{%
{\protect \APACyear {2019}}%
}]{%
stolc2019}
\APACinsertmetastar {%
stolc2019}%
\begin{APACrefauthors}%
{\v{S}tolc}, M.%
\end{APACrefauthors}%
\unskip\
\newblock
\APACrefYearMonthDay{2019}{}{},
\newblock
{\BBOQ}\APACrefatitle {{Accretion Discs in the Context of Tidal Disruption of
  Stars in Nuclei of Galaxies; Master Thesis}} {{Accretion Discs in the Context
  of Tidal Disruption of Stars in Nuclei of Galaxies; Master Thesis}}.{\BBCQ}
\newblock
\BIn{} \APACrefbtitle {(Charles University, Faculty of Mathematics and Physics,
  Prague).} {(Charles University, Faculty of Mathematics and Physics, Prague).}
\PrintBackRefs{\CurrentBib}

\bibitem [\protect \citeauthoryear {%
{Stone}%
}{%
{Stone}%
}{%
{\protect \APACyear {2015}}%
}]{%
Stone}
\APACinsertmetastar {%
Stone}%
\begin{APACrefauthors}%
{Stone}, N\BPBI C.%
\end{APACrefauthors}%
\unskip\
\newblock
\APACrefYearMonthDay{2015}{}{},
\newblock
{\BBOQ}\APACrefatitle {{The Tidal Disruption of Stars by Supermassive Black
  Holes, An Analytic Approach; Ph.D.\ Thesis}} {{The Tidal Disruption of Stars
  by Supermassive Black Holes, An Analytic Approach; Ph.D.\ Thesis}}.{\BBCQ}
\newblock
\BIn{} \APACrefbtitle {(Harvard University, Cambridge, MA).} {(Harvard
  University, Cambridge, MA).}
\PrintBackRefs{\CurrentBib}

\bibitem [\protect \citeauthoryear {%
{Stone}%
, {Kesden}%
, {Cheng}%
\BCBL {}\ \BBA {} {van Velzen}%
}{%
{Stone}%
\ \protect \BOthers {.}}{%
{\protect \APACyear {2019}}%
}]{%
2019GReGr..51...30S}
\APACinsertmetastar {%
2019GReGr..51...30S}%
\begin{APACrefauthors}%
{Stone}, N\BPBI C.%
, {Kesden}, M.%
, {Cheng}, R\BPBI M.%
\BCBL {}\ \BBA {} {van Velzen}, S.%
\end{APACrefauthors}%
\unskip\
\newblock
\APACrefYearMonthDay{2019}{}{},
\newblock
\unskip
\newblock
\APACjournalVolNumPages{General Relativity and Gravitation}{51}{2}{30}.
\newblock
\begin{APACrefURL} \url{https://ui.adsabs.harvard.edu/abs/2019GReGr..51...30S}
  \end{APACrefURL}
\PrintBackRefs{\CurrentBib}

\bibitem [\protect \citeauthoryear {%
{Zhang}%
, {Yu}%
, {Karas}%
\BCBL {}\ \BBA {} {Dov{\v{c}}iak}%
}{%
{Zhang}%
\ \protect \BOthers {.}}{%
{\protect \APACyear {2015}}%
}]{%
2015ApJ...807...89Z}
\APACinsertmetastar {%
2015ApJ...807...89Z}%
\begin{APACrefauthors}%
{Zhang}, W.%
, {Yu}, W.%
, {Karas}, V.%
\BCBL {}\ \BBA {} {Dov{\v{c}}iak}, M.%
\end{APACrefauthors}%
\unskip\
\newblock
\APACrefYearMonthDay{2015}{}{},
\newblock
\unskip
\newblock
\APACjournalVolNumPages{\apj}{807}{1}{89}.
\newblock
\begin{APACrefURL} \url{https://ui.adsabs.harvard.edu/abs/2015ApJ...807...89Z}
  \end{APACrefURL}
\PrintBackRefs{\CurrentBib}

\end{thebibliography}

\end{document}